\RequirePackage{fixltx2e}
\documentclass[twocolumn,prl,floatfix,showpacs,preprintnumbers,nofootinbib,%
superscriptaddress]{revtex4}
\usepackage[english]{babel}
\usepackage{mathrsfs}
\usepackage{amssymb}
\usepackage{amsmath}
\usepackage{mathtools}
\usepackage{mathrsfs}
\usepackage{graphicx}
\usepackage{fixltx2e}

\usepackage{xcolor}
\definecolor{lcolor}{rgb}{0.5,0,0}
\definecolor{citcolor}{rgb}{0,0.3,0.0}

\usepackage[breaklinks,colorlinks,urlcolor=blue,citecolor=citcolor,linkcolor=lcolor]{hyperref}

\newcommand{\der}{\mathrm{d}}
\newcommand{\rt}{{\mathbf{r}_T}}

\newcommand{\bt}{{\mathbf{b}_T}}

\newcommand{\Deltat}{{\boldsymbol{\Delta}_T}}

\newcommand{\ud}{\, \mathrm{d}}

\newcommand{\nr}[1]{(\ref{#1})}

\newcommand{\ra}{R_A}

\newcommand{\qs}{Q_\mathrm{s}}

\newcommand{\eq}{Eq.~}

\newcommand{\xpom}{{x_\mathbb{P}}}

\newcommand{\Aavg}[1]{\left\langle #1 \right\rangle_\textrm{N}}

\newcommand{\A}{{\mathcal{A}}}

\begin{document}

\author{T. Lappi}
\affiliation{
Department of Physics, University of Jyv\"askyl\"a,  P.O. Box 35, 40014 University of Jyv\"askyl\"a, Finland
}

\affiliation{
Helsinki Institute of Physics, P.O. Box 64, 00014 University of Helsinki,
Finland
}

\author{H. M\"antysaari}
\affiliation{
Department of Physics, University of Jyv\"askyl\"a, %
 P.O. Box 35, 40014 University of Jyv\"askyl\"a, Finland
}

\author{R. Venugopalan}
\affiliation{Bldg. 510A, Physics Department, Brookhaven National Laboratory, Upton, NY 11973, USA}

\title{Ballistic protons in incoherent exclusive vector meson production as a measure of rare parton fluctuations at an Electron-Ion Collider
}

\pacs{13.60.-r,24.85.+p}

\preprint{}

\begin{abstract}
We argue that the proton multiplicities measured in Roman pot detectors at an electron ion collider can be used to determine centrality classes in incoherent diffractive scattering. Incoherent diffraction probes the fluctuations in the interaction strengths of multi-parton Fock states in the nuclear wavefunctions.  In particular, the saturation scale that characterizes this multi-parton dynamics is significantly larger in central events relative to minimum bias events. As an application, we study the centrality dependence of incoherent diffractive vector meson production. We identify an observable which is simultaneously very sensitive to centrality triggered parton fluctuations and insensitive to details of the model.
\end{abstract}

\maketitle

\paragraph{Introduction}

Very high multiplicity events in proton-proton (p+p) and proton/deuteron-nucleus (p/d+A) collisions at LHC and RHIC have revealed that the structure of such events is more complex and interesting than previously imagined~\cite{Chatrchyan:2013nka,Abelev:2012ola,Adare:2013piz,Aad:2013fja}. In particular, interpreting the results of these experiments requires a deeper understanding of event-by-event multi-parton spatial fluctuations in protons and 
nuclei~\cite{Schenke:2014zha,Schlichting:2014ipa,Dumitru:2014yza,Werner:2013ipa,Bzdak:2013raa}. Incoherent diffraction in  deeply inelastic scattering (DIS)  of electrons off nuclei (e+A collisions) has been long understood as having the potential to provide insight into event-by-event fluctuations in the spatial structure of nuclei. A significant advantage of e+A collisions relative to p+A collisions is that the former is insensitive to the final state interactions that, in the latter, can complicate the extraction of the spatial parton structure of the proton and the nucleus. 

Insight into rare spatial configurations can be provided by triggering on central incoherent diffractive events in e+A collisions. In diffractive events, no net color charge is exchanged between the fragmentation region of the nucleus and that of the electromagnetic current exciting the nucleus: a rapidity gap is formed between the two fragmentation regions. Coherent diffraction corresponds to the case where the nucleus remains fully intact; in incoherent diffraction,  the $p_T$ kick given to the nucleus is large enough to  break it up, but the rapidity gap is preserved. While the coherent cross section measures the average spatial
distribution of gluons incoherent scattering probes the
fluctuations~\cite{Miettinen:1978jb} and correlations~\cite{Caldwell:2009ke}
in the gluon density.

For incoherent diffractive events in a collider geometry, such as at a future Electron-Ion Collider (EIC) facility~\cite{Accardi:2012qut}, one can distinguish between so-called ballistic nucleons and evaporation nucleons. Ballistic nucleons are produced when a nucleon in the nucleus receives a large longitudinal/transverse momentum kick from the projectile. This nucleon can scatter off other nucleons in the nucleus on its path out. Evaporation nucleons, on the other hand, are produced when the nucleus is excited as a whole, causing it to evaporate nucleons according to a thermal spectrum in the rest frame of the nucleus.

In this work, we will argue that ballistic protons could be used experimentally as a measure of centrality in incoherent diffractive e+A collisions. 
These, unlike evaporation nucleons (or ballistic neutrons), can be measured in forward ``Roman pot'' detectors located in the beam pipe outside the main detector. 
Among those events that have ballistic nucleons (which can be both peripheral or central), the number of ballistic protons should be larger if the nucleus was hit at a central impact parameter.
Since we expect the saturation scale $\qs$ in nuclei in central events to be enhanced relative to minimum bias events, this opens up the possibility to select large $\qs$ events in nuclear DIS by measuring exclusive final states in the central detector in coincidence with recoil protons from the nucleus in the Roman pots. Further, the dependence of the results on kinematic invariants in the scattering shows distinct patterns that make these triggered measurements a sensitive test of the multi-parton dynamics of gluon saturation. 

\begin{figure}[tb]
\centering
		\includegraphics[width=0.4\textwidth]{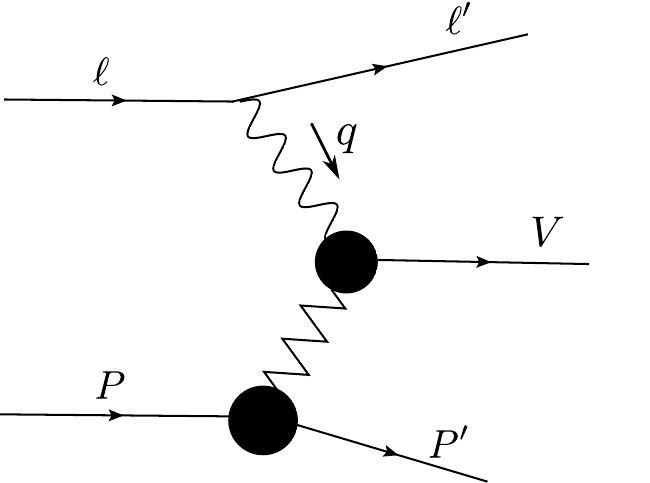} 
		\caption{Diffractive DIS kinematics.}
		\label{fig:ddis}
\end{figure}

\paragraph{Kinematics of diffraction at an EIC}

We will consider the DIS process $e(\ell) + A(P) \rightarrow e(\ell') + A'(P') + J/\Psi(V)$ 
(see Fig. \ref{fig:ddis}). 
Denoting the nucleon/nuclear/vector meson mass by $m_N/m_A/m_V$, the kinematic invariants needed in the process can be expressed as 
\begin{align}
	q^2 &\equiv -Q^2 \equiv (\ell - \ell')^2 \\
	t &\equiv (P'-P)^2 \\
	W^2 &\equiv (P+q)^2 \\
	\xpom &\equiv A \frac{(P-P')\cdot q}{P\cdot q} = A\frac{m_V^2 + Q^2 -t}{W^2+Q^2-m_A^2}  \label{eq:xpom} \\
	\nu &\equiv \frac{P \cdot q}{m_A} \approx \frac{P \cdot q}{A m_N} .
\end{align}
All of these invariants can be determined experimentally by measuring the scattered electron and vector meson four-momenta, even without measuring the recoil nucleus. Note that here $P$ is the momentum of the whole nucleus, thus there is an explicit $A$ in the definition of $\xpom$. The variable $W$ often used to describe DIS off protons at HERA is  not very natural for nuclei since $W^2 = A^2 m_N^2 + 2 Am_N \nu - Q^2$ does not scale in a simple way with $A$ at fixed beam energies per nucleon.

To be specific, we will consider an EIC with 15 GeV electron beams scattering off nuclear beams with 100 GeV/nucleon. We need to first establish the optimal kinematics for our study. We need a small $\xpom \lesssim 0.01$, for saturation effects to be relevant and to have a significant rapidity gap ($\sim \ln(1/\xpom)$) between the current and target fragmentation regions.
 The EIC energy therefore effectively restricts us to $Q^2 \lesssim 10$ GeV$^2$.  Coherent diffraction dominates the exclusive cross section at low values of $|t| \sim 1/R_A^2$, 
dying very rapidly. Incoherent diffraction off the nucleus dominates in the kinematic regime $1/R_A^2 \lesssim |t| \lesssim 1/R_p^2$, where $R_p$ is the nucleon radius. For $|t| \gtrsim 1/R_p^2$, incoherent diffraction will be sensitive to sub-nucleon scale fluctuations in the nucleus.
 Because of the small momentum transfer to the target, coherent scattering probes the whole transverse plane of the nucleus, and it is not possible to identify a well
defined impact parameter event-by-event. For incoherent scattering, on the other hand, the scattering is localized to an area $\sim R_p^2$ in the transverse plane, and one can classify individual events into centrality classes.

\begin{figure}[tb]
\centering
		\includegraphics[width=0.5\textwidth]{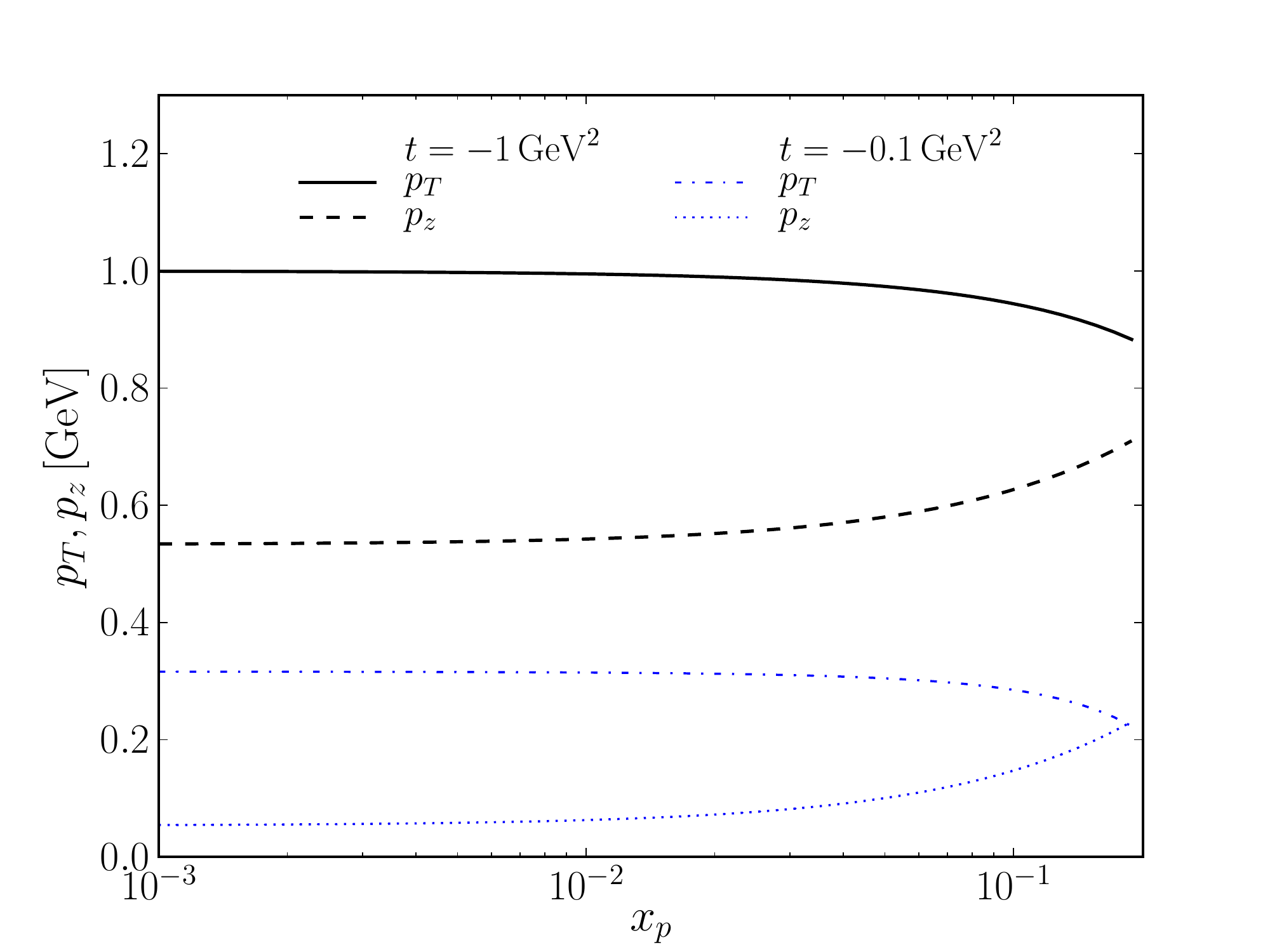} 
		\caption{Scattered proton longitudinal and transverse momentum in the target rest frame in diffractive J/$\Psi$ production. Here we choose $Q^2=1$ GeV$^2$.}
		\label{fig:TRF}
\end{figure}

It is interesting to consider what the collision looks like in the target rest frame (TRF). The TRF is defined as the frame where the nucleus four-momentum is $P=(A m_N,0,0,0)$ and the $z$ axis is defined along the direction of the photon momentum: $q=(\nu,0_T,\sqrt{\nu^2+Q^2})$. The $\gamma^\star p$ scattering kinematics is then fixed by the three invariants $\xpom$, $Q^2$ and $t$.  In Fig.~\ref{fig:TRF}, we show the dependence of the longitudinal and transverse momenta of the scattered proton, $p_z, p_T$, defined relative to the photon $z$ axis, as a function of $\xpom$, for $t=-0.1$ and $-1$ GeV$^2$, encompassing the impact parameter range between sub-nuclear to sub-nucleon scale fluctuations. Since (neglecting terms $\sim Q^2/\nu^2$), the momentum transfer is $t = -(m_N^2 \xpom^2 +(p_T^{\rm TRF})^2)/(1-\xpom/A)$, one has $t\approx -(p_T^{\rm TRF})^2$ for a wide range in $\xpom$. 
The exact relation between the recoil transverse momenta in the TRF and the collider frames depends on the lepton kinematics--these momenta are however quite close to each other at high energies\footnote{The angle between the incoming lepton and the $\gamma^*$ is small in the TRF. Since the outgoing recoil longitudinal momentum in the TRF is small, the rotation by this angle from the $\gamma^*$ TRF to the electron TRF does not change the recoil $p_T$ much. The further boost to the collider 
frame clearly has no effect on $p_T$. Hence, $p_T^{\rm TRF}\approx p_T^{\rm Collider}$.}.

In the longitudinal direction, the coherence length at small $\xpom$ is large. As noted previously, however,  in the transverse plane momentum is first deposited in a nucleon-size area with a well defined impact parameter.  The struck nucleon, or nucleons, can then rescatter on their paths out of the nucleus. The process whereby this occurs is complex and can result in the break-up of the nucleus into fragments, leading at later times to evaporation of nucleons from the fragments. For a discussion of this dynamics, see for instance \cite{Aichelin:1986zz,Hirenzaki:1995js} and references therein or 
Refs~\cite{Baltz:2002pp,Rebyakova:2011vf,Guzey:2013jaa} for a different
approach.  
Despite this complexity, a relatively clean separation exists, of over an order of magnitude,  in the typical transverse momentum scales of ballistic nucleons (with a sizeable fraction of the original momentum transfer) and evaporation nucleons. Nucleons with laboratory transverse momentum in the $400$~MeV\dots$1$~GeV range of interest can be clearly identified as being ballistic.  Since the energy and momentum transport of such nucleons is likely well localized, the measured multiplicity of the latter will be a sensitive trigger of centrality in diffractive final states. This is in contrast to evaporative nuclear breakup alone where all information about the initial impact parameter is lost due to the thermalization of the excited nucleus.

\begin{figure}[tb]
\centering
		\includegraphics[width=0.5\textwidth]{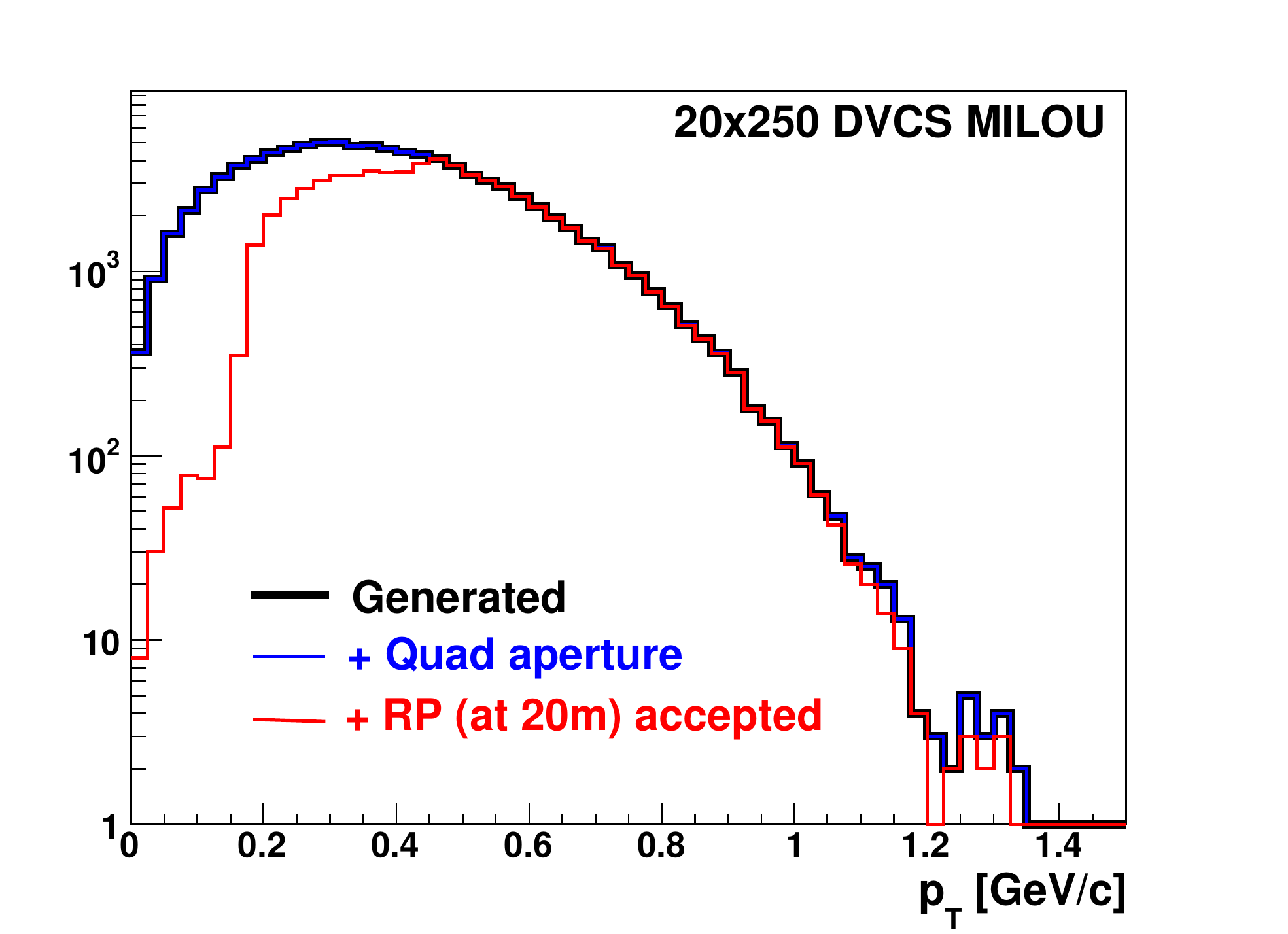} 
		\caption{
EIC acceptance for protons, as a function of $p_T$, in a Roman pot detector. Figure from~\cite{eicwikidvcs}. The reaction simulated is Deeply Virtual 
Compton Scattering (DVCS) in e+p scattering. The region of strong overlap between three curves demarcates the 
$p_T$ acceptance. Because of the different configuration of magnetic fields in e+A scattering, there will be a shift in the 
acceptance to lower $p_T$ than shown here.
}
		\label{fig:eic_dvcs}
\end{figure}

As noted, the multiplicity of ballistic protons can be measured with Roman pots. As an illustration, we show  in   Fig. \ref{fig:eic_dvcs} the results of a simulation for e+p  Deeply Virtual Compton Scattering (DVCS). One observes that the acceptance is excellent in the $p_T$ region of incoherent diffraction in nuclei. The acceptance for e+A scattering will not be exactly the same due to the different magnetic fields required by the nuclear $Z/A$ ratio and will require separate detector simulations to estimate. However, we would still expect good resolution in a sizeable part of the relevant $p_T$ region.
On the other hand, ballistic neutrons, while perhaps measurable in a Zero Degree Calorimeter (ZDC), can be challenging to separate from evaporation neutrons~\cite{Zheng:2014cha}.

\paragraph{Diffractive vector meson production}

As a model example, we will explore the centrality dependence of incoherent diffractive vector meson production in the saturation model of Ref.~\cite{Lappi:2010dd}. In this framework, diffractive scattering is described such that an incoming virtual photon fluctuates into a quark-antiquark color dipole which scatters off the target and forms the final state vector meson. The necessary ingredients in these calculations are the dipole-nucleus scattering amplitude $N_A$ and the vector meson photon wave function overlap $\Psi^*_V\Psi$. The imaginary part of the scattering amplitude for the $\gamma^* A \to V A$ scattering is
\begin{multline}
\label{eq:diffr-amp}
\A(\xpom,Q^2,\Deltat) 
= \int \ud^2 \rt \int \frac{\ud z}{4\pi} \int \ud^2 \bt 
\\
\times [\Psi_V^* \Psi](r,Q^2,z)
e^{-i \bt \cdot  \Deltat}  
2\, N_A(\rt, \bt, \xpom),
\end{multline}
where $\bt$ is the impact parameter, $\rt$ the dipole size, $\Deltat$ the  momentum transfer, $t=-\Deltat^2$ and $z$ is the longitudinal momentum fraction of the photon carried by the quark. 

The dipole-nucleus amplitude is obtained from the dipole-proton amplitude $N_p$ by taking the independent scattering approximation and writing 
$S=1-N$ as
\begin{equation}
	S_A(\rt,\bt,\xpom) = \prod_{i=1}^A S_p(\rt,\bt-\mathbf{b}_{Ti},\xpom),
\end{equation}
where $\mathbf{b}_{Ti}$ are nucleon coordinates.
For the dipole-proton amplitude $N_p$ we use the IPsat model~\cite{Kowalski:2003hm} which has an eikonalized DGLAP-evolved gluon distribution. The model parameters are fit to the HERA data in Ref.~\cite{Kowalski:2006hc}.
To simplify the calculation of the incoherent cross section we introduce a 
factorized approximation
%
\begin{equation}
	S_p(\rt,\bt,\xpom) = 1 - T_p(\bt) N_p(\rt,\xpom),
\end{equation}
with a Gaussian nucleon profile $T_p(\bt)$.
%
%
For the vector meson overlaps
$\Psi^*_V\Psi$ we use the boosted Gaussian parametrization from Ref.~\cite{Kowalski:2006hc}.

The cross section for coherent diffractive vector meson production for a scattering off a nucleus can be calculated by averaging the scattering amplitude $\A$ over the nucleon configurations and then taking the square:
\begin{equation} \label{eq:xsec}
\frac{\der \sigma^{\gamma^* A \to V A }}{\der t} 
= \frac{1}{16\pi} \left|\Aavg{\A(\xpom,Q^2,\Deltat)}\right|^2,
\end{equation}
where the average is defined as
\begin{equation} \label{eq:aavg}
\Aavg{\mathcal{O}(\{ \bt_i \})} 
\equiv \int \prod_{i=1}^{A}\left[ \ud^2 \bt_i T_A(\bt_i) \right] 
\mathcal{O}(\{ \bt_i \})
\end{equation}
and $T_A$ refers to the Woods-Saxon distribution. The incoherent cross section is similarly given by the variance $\Aavg{|\A|^2}-|\Aavg{\A}|^2$. Full expressions for the coherent and incoherent diffractive cross sections can be found from Ref.~\cite{Lappi:2010dd} (see also Refs.~\cite{Kopeliovich:2001xj,Kowalski:2008sa,Toll:2012mb,Santos:2014vwa}).

This framework gives a good description of the diffractive vector meson production in electron-proton scattering measured at HERA--see Refs.~\cite{Kowalski:2006hc,Rezaeian:2012ji}. 
Exclusive photon-nucleus collisions have also been studied in  ultraperipheral heavy ion collisions by the  at both 
RHIC~\cite{Adler:2002sc,Afanasiev:2009hy} 
and LHC~\cite{Abbas:2013oua,Abelev:2012ba}.
The ALICE results 
for $J/\Psi$ production~\cite{Abbas:2013oua,Abelev:2012ba}
are consistent with the calculations of Ref.~\cite{Lappi:2013am} using the dipole model of Ref.~\cite{Lappi:2010dd}.

\begin{figure}[tb]
\centering
		\includegraphics[width=0.52\textwidth]{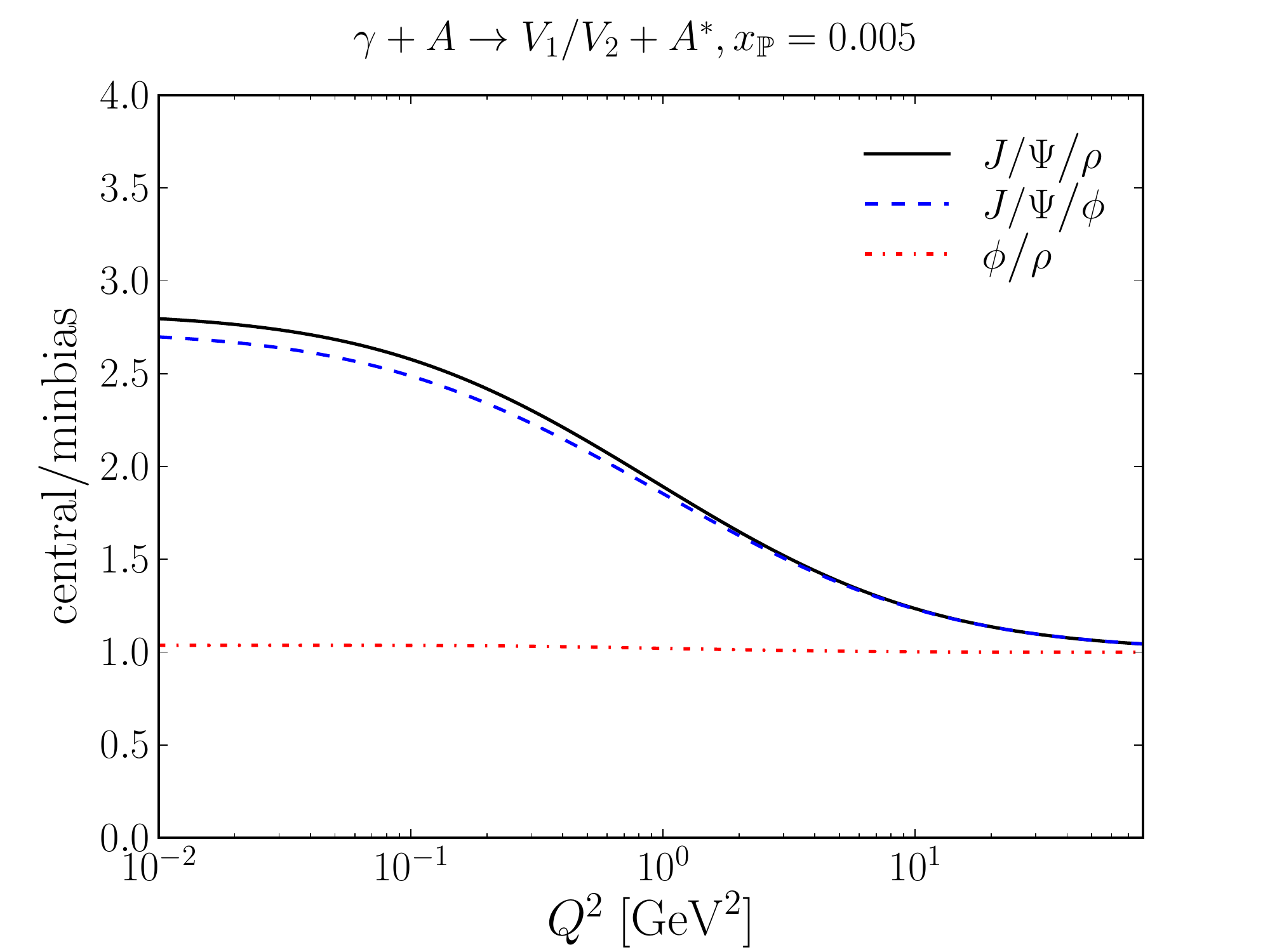} 
		\caption{Ratio of two vector meson incoherent diffractive cross-sections in central events relative to minimum bias events as a function of $Q^2$. See text for details.
}
		\label{fig:central-minbias}
\end{figure}

We compute diffractive vector meson production in two centrality classes. 
For simplicity ``central'' events are calculated at $b=0$ but the event characteristics are not expected to depend strongly on $b$ for $b\ll \ra$.  As discussed previously, these are the events that should have large proton multiplicities in the Roman pot detectors. These events are compared with minimum bias results obtained by integrating over all impact 
parameters\footnote{We do not have a reliable estimate of the share of ballistic events among all incoherent events; therefore we cannot calculate a ballistic event cross section. This factor, however, cancels in the double cross section ratio discussed in the following.}.

In Fig.~\ref{fig:central-minbias}, we show the $Q^2$ dependence of ratios of incoherent diffractive production cross sections of different vector mesons off a gold nucleus in these central events relative to those in minimum bias,
\begin{equation}
	\label{eq:doubleratio}
	 \frac{ \left. \sigma(\gamma^*A \to V_1 A^*) \Big/  \sigma(\gamma^*A \to V_2 A^*) \right|_{\text{central}} }
	 {\left. \sigma(\gamma^*A \to V_1 A^*) \Big/  \sigma(\gamma^* A \to V_2 A^*) \right|_{\text{minimum bias}} },
\end{equation}
where $V_1$ and $V_2$ are different vector mesons (e.g. J/$\Psi$ and $\rho$) and $A^*$ refers to the nucleus that breaks up.

The double ratio is useful because both uncertainties in the overall normalizations and those in the vector meson wavefunctions are minimized by taking this ratio. The calculation is done at $\xpom = 0.005$ 
 -- this is well within the EIC kinematic reach.
We have checked that the $\xpom$ and $W$ dependence of the results is quite weak. Further, in the approximations employed here, this ratio is independent of $t$  and contributions from the real part of the amplitude and skewness corrections cancel in the double ratio.

We see from Fig.~\ref{fig:central-minbias} that a very significant enhancement for the $J/\Psi/\rho$ and $J/\Psi/\phi$ double ratio is seen at low $Q^2$; it reduces to unity only above $Q^2 \sim 10$ GeV$^2$ when $Q^2$ becomes much larger than the central and minimum bias saturation scales. In contrast, the $\phi/\rho$ remains nearly at unity for the entire range in $Q^2$ studied. 

The result has the following simple interpretation. Let's first consider the $J/\Psi/\rho$ and $J/\Psi/\phi$ ratios. The small size of the $J/\Psi$ wavefunction in Eq.~\ref{eq:diffr-amp} indicates that the dipole amplitude is dominated by ``color transparent" small size configurations with $r^2\qs^2 \ll 1$, even at low $Q^2$. Hence the production cross-section for incoherent diffractive $J/\Psi$ goes as $\qs^4$.  In contrast, since the $\rho$ and $\phi$ meson wavefunctions are significantly broader, the corresponding typical configurations in that case have $r^2\qs^2\geq 1$ in both central and minimum bias events. These ``color opaque"  $\rho$ and $\phi$ configurations have cross-sections of the order of the geometrical radius for both central and minimum bias events and these cancel in the double ratio. Hence 
\eq\nr{eq:doubleratio} is the central-to-minimum-bias ratio of the color transparent $J/\Psi$ cross-sections, which goes as ${\qs}_{\rm central}^4/{\qs}_{\rm min.bias}^4$. At large $Q^2$, even the $\phi$ and $\rho$ cross-sections become 
color transparent. Thus the saturation scale in both the numerator and the denominator cancel separately, and one obtains unity as seen in Fig.~\ref{fig:central-minbias}. 

In contrast, since the $\phi$ and the $\rho$ are simultaneously either color opaque or color transparent depending on the $Q^2$ probed, there is never a strong sensitivity to $\qs^2$ and the double ratio is close to unity for all $Q^2$. We emphasize that the saturation  scale $\qs$ should really be thought of as a transverse momentum/length scale, and its cleanest manifestations in DIS should be in the $Q^2$ dependence of observables. Therefore the result shown in Fig.~\ref{fig:central-minbias} is a clear direct measure of the nuclear enhancement of nonlinear gluon dynamics, the large  nuclear ``oomph'', previously quantified for inclusive DIS nuclear cross-sections relative to inclusive proton cross-sections~\cite{Kowalski:2007rw}.

\paragraph{Discussion}

We argued that ``ballistic protons'' can be used as a measure of centrality in diffractive processes at an electron ion collider by measuring proton multiplicities in the Roman pot detectors.
Triggering on the highest multiplicity (most central) events makes it possible to probe fluctuations in the interaction strengths of rare parton configurations (with large $\qs$) in the nuclear wave function at high energies. As an example, we showed that the double ratio of the production cross sections of different vector meson species in central and minimum bias collisions has a large $Q^2$ dependence. We anticipate this double ratio will be significantly different in models where fluctuations in the parton Fock state configurations are treated differently than in dipole models. 
 It would be interesting to combine model calculations for the production cross section with a more detailed model for the nuclear breakup.

Centrality selection in nuclear DIS using the multiplicity of evaporation neutrons measured in the  ZDC was recently discussed in Ref.~\cite{Zheng:2014cha}. A potential impact of this centrality selection on single inclusive multiplicities and di-hadron correlations was also discussed. This study was performed in the context of inclusive scattering, where the nuclear excitation and breakup can be very different than for diffraction considered here. This approach might provide complementary information on the dynamics of rare large-$\qs$ configurations in high energy QCD.

\section*{Acknowledgements} 
We thank E. Aschenauer for discussions and useful comments on the manuscript.
H.M. and T.L. are supported by the Academy of Finland, projects 
267321 and 273464, and the Graduate School of Particle and Nuclear Physics (H.M.). H.M. wishes to thank the nuclear theory group at BNL for hospitality during the early stages of this work. R.V.'s work is supported by the US Department of Energy under DOE Contract No.~DE-AC02-98CH10886.

\bibliographystyle{../bib/JHEP-2mod}
\bibliography{../bib/spires}

\providecommand{\href}[2]{#2}\begingroup\raggedright\begin{thebibliography}{10}

\bibitem{Chatrchyan:2013nka}
{\bf CMS} collaboration, S.~Chatrchyan {\em et.~al.},  \mbox{}
  \href{http://dx.doi.org/10.1016/j.physletb.2013.06.028}{{\em Phys. Lett.}
  {\bf B724} (2013) 213} [\href{http://arXiv.org/abs/1305.0609}{{\tt
  arXiv:1305.0609 [nucl-ex]}}].

\bibitem{Abelev:2012ola}
{\bf ALICE} collaboration, B.~Abelev {\em et.~al.},  \mbox{}
  \href{http://dx.doi.org/10.1016/j.physletb.2013.01.012}{{\em Phys. Lett.}
  {\bf B719} (2013) 29} [\href{http://arXiv.org/abs/1212.2001}{{\tt
  arXiv:1212.2001}}].

\bibitem{Adare:2013piz}
{\bf PHENIX} collaboration, A.~Adare {\em et.~al.},  \mbox{}
  \href{http://dx.doi.org/10.1103/PhysRevLett.111.212301}{{\em Phys. Rev.
  Lett.} {\bf 111} (2013) 212301} [\href{http://arXiv.org/abs/1303.1794}{{\tt
  arXiv:1303.1794 [nucl-ex]}}].

\bibitem{Aad:2013fja}
{\bf ATLAS} collaboration, G.~Aad {\em et.~al.},  \mbox{}
  \href{http://dx.doi.org/10.1016/j.physletb.2013.06.057}{{\em Phys. Lett.}
  {\bf B725} (2013) 60} [\href{http://arXiv.org/abs/1303.2084}{{\tt
  arXiv:1303.2084 [hep-ex]}}].

\bibitem{Schenke:2014zha}
B.~Schenke and R.~Venugopalan,  \mbox{}
  \href{http://dx.doi.org/10.1103/PhysRevLett.113.102301}{{\em Phys. Rev.
  Lett.} {\bf 113} (2014) 102301} [\href{http://arXiv.org/abs/1405.3605}{{\tt
  arXiv:1405.3605 [nucl-th]}}].

\bibitem{Schlichting:2014ipa}
S.~Schlichting and B.~Schenke,  \mbox{}
  \href{http://dx.doi.org/10.1016/j.physletb.2014.10.068}{{\em Phys. Lett.}
  {\bf B739} (2014) 313} [\href{http://arXiv.org/abs/1407.8458}{{\tt
  arXiv:1407.8458 [hep-ph]}}].

\bibitem{Dumitru:2014yza}
A.~Dumitru, L.~McLerran and V.~Skokov,  \mbox{}
  \href{http://arXiv.org/abs/1410.4844}{{\tt arXiv:1410.4844 [hep-ph]}}.

\bibitem{Werner:2013ipa}
K.~Werner, M.~Bleicher, B.~Guiot, I.~Karpenko and T.~Pierog,  \mbox{}
  \href{http://arXiv.org/abs/1307.4379}{{\tt arXiv:1307.4379 [nucl-th]}}.

\bibitem{Bzdak:2013raa}
A.~Bzdak and V.~Skokov,  \mbox{}  \href{http://arXiv.org/abs/1312.7349}{{\tt
  arXiv:1312.7349 [hep-ph]}}.

\bibitem{Miettinen:1978jb}
H.~I. Miettinen and J.~Pumplin,  \mbox{}
  \href{http://dx.doi.org/10.1103/PhysRevD.18.1696}{{\em Phys. Rev.} {\bf D18}
  (1978) 1696}.

\bibitem{Caldwell:2009ke}
A.~Caldwell and H.~Kowalski,  \mbox{}
  \href{http://dx.doi.org/10.1103/PhysRevC.81.025203}{{\em Phys. Rev.} {\bf
  C81} (2010) 025203} [\href{http://arXiv.org/abs/0909.1254}{{\tt
  arXiv:0909.1254 [hep-ph]}}].

\bibitem{Accardi:2012qut}
A.~Accardi, J.~Albacete, M.~Anselmino, N.~Armesto, E.~Aschenauer {\em et.~al.},
   \mbox{}  \href{http://arXiv.org/abs/1212.1701}{{\tt arXiv:1212.1701
  [nucl-ex]}}.

\bibitem{Aichelin:1986zz}
J.~Aichelin,  \mbox{}  \href{http://dx.doi.org/10.1103/PhysRevC.33.537}{{\em
  Phys. Rev.} {\bf C33} (1986) 537}.

\bibitem{Hirenzaki:1995js}
S.~Hirenzaki, P.~Fernandez~de Cordoba and E.~Oset,  \mbox{}
  \href{http://dx.doi.org/10.1103/PhysRevC.53.277}{{\em Phys. Rev.} {\bf C53}
  (1996) 277} [\href{http://arXiv.org/abs/nucl-th/9511036}{{\tt
  arXiv:nucl-th/9511036 [nucl-th]}}].

\bibitem{Baltz:2002pp}
A.~J. Baltz, S.~R. Klein and J.~Nystrand,  \mbox{}
  \href{http://dx.doi.org/10.1103/PhysRevLett.89.012301}{{\em Phys. Rev. Lett.}
  {\bf 89} (2002) 012301} [\href{http://arXiv.org/abs/nucl-th/0205031}{{\tt
  arXiv:nucl-th/0205031 [nucl-th]}}].

\bibitem{Rebyakova:2011vf}
V.~Rebyakova, M.~Strikman and M.~Zhalov,  \mbox{}
  \href{http://dx.doi.org/10.1016/j.physletb.2012.03.041}{{\em Phys. Lett.}
  {\bf B710} (2012) 647} [\href{http://arXiv.org/abs/1109.0737}{{\tt
  arXiv:1109.0737 [hep-ph]}}].

\bibitem{Guzey:2013jaa}
V.~Guzey, M.~Strikman and M.~Zhalov,  \mbox{}
  \href{http://dx.doi.org/10.1140/epjc/s10052-014-2942-z}{{\em Eur. Phys. J.}
  {\bf C74} (2014) 2942} [\href{http://arXiv.org/abs/1312.6486}{{\tt
  arXiv:1312.6486 [hep-ph]}}].

\bibitem{Zheng:2014cha}
L.~Zheng, E.~Aschenauer and J.~Lee,  \mbox{}
  \href{http://arXiv.org/abs/1407.8055}{{\tt arXiv:1407.8055 [hep-ex]}}.

\bibitem{eicwikidvcs}
\url{https://wiki.bnl.gov/eic/index.php/Detector_Design_Requirements}.

\bibitem{Lappi:2010dd}
T.~Lappi and H.~M{\"a}ntysaari,  \mbox{}
  \href{http://dx.doi.org/10.1103/PhysRevC.83.065202}{{\em Phys. Rev.} {\bf
  C83} (2011) 065202} [\href{http://arXiv.org/abs/1011.1988}{{\tt
  arXiv:1011.1988 [hep-ph]}}].

\bibitem{Kowalski:2003hm}
H.~Kowalski and D.~Teaney,  \mbox{}
  \href{http://dx.doi.org/10.1103/PhysRevD.68.114005}{{\em Phys. Rev.} {\bf
  D68} (2003) 114005} [\href{http://arXiv.org/abs/hep-ph/0304189}{{\tt
  arXiv:hep-ph/0304189}}].

\bibitem{Kowalski:2006hc}
H.~Kowalski, L.~Motyka and G.~Watt,  \mbox{}
  \href{http://dx.doi.org/10.1103/PhysRevD.74.074016}{{\em Phys. Rev.} {\bf
  D74} (2006) 074016} [\href{http://arXiv.org/abs/hep-ph/0606272}{{\tt
  arXiv:hep-ph/0606272}}].

\bibitem{Kopeliovich:2001xj}
B.~Z. Kopeliovich, J.~Nemchik, A.~Schafer and A.~V. Tarasov,  \mbox{}
  \href{http://dx.doi.org/10.1103/PhysRevC.65.035201}{{\em Phys. Rev.} {\bf
  C65} (2002) 035201} [\href{http://arXiv.org/abs/hep-ph/0107227}{{\tt
  arXiv:hep-ph/0107227}}].

\bibitem{Kowalski:2008sa}
H.~Kowalski, T.~Lappi, C.~Marquet and R.~Venugopalan,  \mbox{}
  \href{http://dx.doi.org/10.1103/PhysRevC.78.045201}{{\em Phys. Rev.} {\bf
  C78} (2008) 045201} [\href{http://arXiv.org/abs/0805.4071}{{\tt
  arXiv:0805.4071 [hep-ph]}}].

\bibitem{Toll:2012mb}
T.~Toll and T.~Ullrich,  \mbox{}
  \href{http://dx.doi.org/10.1103/PhysRevC.87.024913}{{\em Phys. Rev.} {\bf
  C87} (2013) 024913} [\href{http://arXiv.org/abs/1211.3048}{{\tt
  arXiv:1211.3048 [hep-ph]}}].

\bibitem{Santos:2014vwa}
G.~S. dos Santos and M.~Machado,  \mbox{}
  \href{http://arXiv.org/abs/1407.4148}{{\tt arXiv:1407.4148 [hep-ph]}}.

\bibitem{Rezaeian:2012ji}
A.~H. Rezaeian, M.~Siddikov, M.~Van~de Klundert and R.~Venugopalan,  \mbox{}
  \href{http://dx.doi.org/10.1103/PhysRevD.87.034002}{{\em Phys. Rev.} {\bf
  D87} (2013) 034002} [\href{http://arXiv.org/abs/1212.2974}{{\tt
  arXiv:1212.2974 [hep-ph]}}].

\bibitem{Adler:2002sc}
{\bf STAR} collaboration, C.~Adler {\em et.~al.},  \mbox{}
  \href{http://dx.doi.org/10.1103/PhysRevLett.89.272302}{{\em Phys. Rev. Lett.}
  {\bf 89} (2002) 272302} [\href{http://arXiv.org/abs/nucl-ex/0206004}{{\tt
  arXiv:nucl-ex/0206004 [nucl-ex]}}].

\bibitem{Afanasiev:2009hy}
{\bf PHENIX} collaboration, S.~Afanasiev {\em et.~al.},  \mbox{}
  \href{http://dx.doi.org/10.1016/j.physletb.2009.07.061}{{\em Phys. Lett.}
  {\bf B679} (2009) 321} [\href{http://arXiv.org/abs/0903.2041}{{\tt
  arXiv:0903.2041 [nucl-ex]}}].

\bibitem{Abbas:2013oua}
{\bf ALICE} collaboration, E.~Abbas {\em et.~al.},  \mbox{}
  \href{http://dx.doi.org/10.1140/epjc/s10052-013-2617-1}{{\em Eur. Phys. J.}
  {\bf C73} (2013)no.~11 2617} [\href{http://arXiv.org/abs/1305.1467}{{\tt
  arXiv:1305.1467 [nucl-ex]}}].

\bibitem{Abelev:2012ba}
{\bf ALICE} collaboration, B.~Abelev {\em et.~al.},  \mbox{}
  \href{http://dx.doi.org/10.1016/j.physletb.2012.11.059}{{\em Phys. Lett.}
  {\bf B718} (2013) 1273} [\href{http://arXiv.org/abs/1209.3715}{{\tt
  arXiv:1209.3715 [nucl-ex]}}].

\bibitem{Lappi:2013am}
T.~Lappi and H.~M{\"a}ntysaari,  \mbox{}
  \href{http://dx.doi.org/10.1103/PhysRevC.87.032201}{{\em Phys. Rev.} {\bf
  C87} (2013) 032201} [\href{http://arXiv.org/abs/1301.4095}{{\tt
  arXiv:1301.4095 [hep-ph]}}].

\bibitem{Kowalski:2007rw}
H.~Kowalski, T.~Lappi and R.~Venugopalan,  \mbox{}
  \href{http://dx.doi.org/10.1103/PhysRevLett.100.022303}{{\em Phys. Rev.
  Lett.} {\bf 100} (2008) 022303} [\href{http://arXiv.org/abs/0705.3047}{{\tt
  arXiv:0705.3047 [hep-ph]}}].

\end{thebibliography}\endgroup

\end{document}